\documentclass[conference]{IEEEtran}
\usepackage[latin9]{inputenc}
\usepackage{textcomp}
\usepackage{amsmath}
\usepackage{amssymb}
\usepackage{graphicx}

\makeatletter

\newcommand*\LyXThinSpace{\,\hspace{0pt}}
\DeclareFontEncoding{LGR}{}{}
\DeclareRobustCommand{\greektext}{%
  \fontencoding{LGR}\selectfont\def\encodingdefault{LGR}}
\DeclareRobustCommand{\textgreek}[1]{\leavevmode{\greektext #1}}
\ProvideTextCommand{\~}{LGR}[1]{\char126#1}

\providecommand{\tabularnewline}{\\}

\newenvironment{lyxlist}[1]
	{\begin{list}{}
		{\settowidth{\labelwidth}{#1}
		 \setlength{\leftmargin}{\labelwidth}
		 \addtolength{\leftmargin}{\labelsep}
		 }}
	{\end{list}}

\IEEEoverridecommandlockouts
\usepackage{cite}
\usepackage{amsfonts}\usepackage{algorithmic}
\usepackage{xcolor}
\def\BibTeX{{\rm B\kern-.05em{\sc i\kern-.025em b}\kern-.08em
    T\kern-.1667em\lower.7ex\hbox{E}\kern-.125emX}}

\makeatother

\begin{document}
\title{Machine Learning for Performance-Aware Virtual Network Function Placement}
\author{\IEEEauthorblockN{Dimitrios~Michael~Manias\IEEEauthorrefmark{1}, Manar~Jammal\IEEEauthorrefmark{1},
Hassan~Hawilo\IEEEauthorrefmark{1}, Abdallah~Shami\IEEEauthorrefmark{1},
Parisa~Heidari\IEEEauthorrefmark{2}, Adel~Larabi\IEEEauthorrefmark{2},
\\
 and Richard~Brunner\IEEEauthorrefmark{2}} ECE Department, Western University, London ON, Canada \IEEEauthorrefmark{1},
Edge Gravity by Ericsson, Montreal, Canada\IEEEauthorrefmark{2}\\
 \{dmanias3, mjammal, hhawilo, Abdallah.shami\}@uwo.ca,\\ \{PHeidari,
ALarabi, RBrunner\}@edgegravity.ericsson.com}
\maketitle
\begin{abstract}
With the growing demand for data connectivity, network service providers
are faced with the task of reducing their capital and operational
expenses while simultaneously improving network performance and addressing
the increased connectivity demand. Although Network Function Virtualization
(NFV) has been identified as a solution, several challenges must be
addressed to ensure its feasibility. In this paper, we address the
Virtual Network Function (VNF) placement problem by developing a machine
learning decision tree model that learns from the effective placement
of the various VNF instances forming a Service Function Chain (SFC).
The model takes several performance-related features from the network
as an input and selects the placement of the various VNF instances
on network servers with the objective of minimizing the delay between
dependent VNF instances. The benefits of using machine learning are
realized by moving away from a complex mathematical modelling of the
system and towards a data-based understanding of the system. Using
the Evolved Packet Core (EPC) as a use case, we evaluate our model
on different data center networks and compare it to the BACON algorithm
in terms of the delay between interconnected components and the total
delay across the SFC. Furthermore, a time complexity analysis is performed
to show the effectiveness of the model in NFV applications. 
\end{abstract}


\section{Introduction}

In recent years, the fast-paced increase in the number of data-producing
connected devices has put an incredible burden on Network Service
Providers (NSPs) worldwide. One of the main challenges currently facing
these NSPs is delivering a continuous and quality service while simultaneously
addressing the increasing connectivity demand. It is estimated that
by the year 2022, the number of connected devices will greatly exceed
the global population by a factor of three \cite{key-1}. Furthermore,
the spike in devices will also contribute to a significant increase
in the amount of Internet Protocol (IP) traffic worldwide \cite{key-1}.

In order to cope with these challenges and the burden on the system,
NSPs need to adapt their networks to enable increased flexibility,
scalability and portability. Network Function Virtualization (NFV)
has been proposed as a solution to these challenges. The goal of NFV
architecture is to isolate the network functions from their underlying
hardware and execute them as software-based applications on servers
and in data centers \cite{key-2}. Several potential benefits arise
from the implementation of NFV architecture including a reduction
in capital and operational expenditures, decreased time to market
for new technologies, service testing and implementation efficiencies,
network topology optimization, optimized energy consumption, and increased
operational efficiencies \cite{key-3}. However, there are challenges
associated with these benefits that must be solved in order to experience
the full potential and power of this technology.

NPSs worldwide are held to certain standards when providing a service
to a customer. Quality of Service (QoS) requirements take into consideration
metrics such as packet loss, jitter, transmission delay, and availability.
The QoS guarantee is an NSPs acknowledgement of and adherence to these
requirements. When considering the implementation of NFV architecture,
QoS guarantees are of paramount importance and consideration. Therefore,
violating these requirements jeopardizes the implementation of NFV
technology in current networks.

Performance, being a key metric of QoS, plays a major role in the
successful implementation of NFV architecture. While the performance
of an individual VNF instance is important, the performance of an
interconnected and interdependent group of VNF instances known as
a Service Function Chain (SFC) is paramount. SFCs are the end goal
of NFV enabled networks. In order to provide an end-to-end service,
several VNF instances of different types should be accessed in a specific
order, thus creating an SFC.

\begin{figure}[h]
\centerline{\includegraphics[width=7.5cm,height=7.5cm]{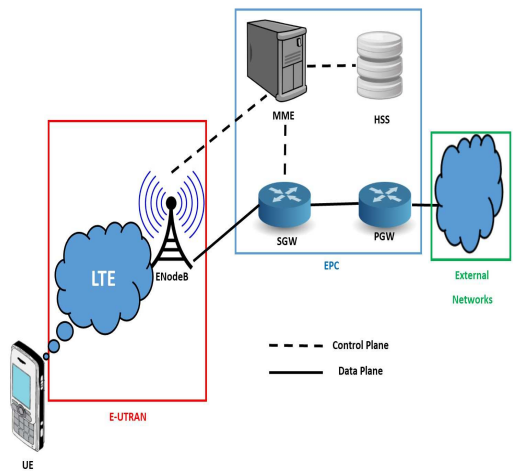}} \caption{Evolved Packet Core (EPC) Architecture}
\label{fig-1} 
\end{figure}

An example of an SFC is the Evolved Packet Core (EPC) which is a network
infrastructure that supports the converging on licensed and unlicensed
radio access technologies through IP \cite{key-4}. Virtual EPC (vEPC)
is a solution introduced by 3GPP to harness the full potential of
radio access technologies \cite{key-5}. In this technology, there
are four main types of VNFs: the Home Subscriber Service (HSS), the
Mobility Management Entity (MME), the Serving Gateway (SGW), and the
Packet Data Network Gateway (PGW). Fig.~\ref{fig-1} outlines the
architecture of this technology. EPC was selected as a use case since
it is one of the candidate network entities for virtualization.

As with any SFC, vEPC is subject to QoS guarantees including performance,
reliability, and availability. When considering a VNF-enable network,
the placement of each of the VNF instances forming the SFC directly
impacts these requirements. When the initial placement of VNF instances
is being conducted, the NFV orchestrator must take into consideration,
among other things, the delay requirements of each SFC when placing
the VNF instances. Furthermore, an intelligent orchestrator may use
machine learning in order to learn from previous placements to predict
the placement of future VNF instances such that the delay between
interdependent components is reduced.

The work outlined in this paper presents the use of machine learning
algorithms in the NFV orchestrator to predict initial VNF instance
placement in a network while taking into consideration QoS guarantees.
Our work is conducted using the EPC infrastructure as a use case and
can generalize to any SFC type. The BACON algorithm is a near-optimal
placement algorithm that considers the minimization of the delay between
dependent VNF instances and across the overall SFC \cite{key-6}.
This algorithm is used to assess the performance of our model. The
results show that the machine learning algorithm can learn from effective
previous placements, QoS requirements, network conditions, and other
operational constraints and translate them into elements of intelligence
to predict placements while minimizing delay.

The remainder of this paper is structured as follows. Section II presents
related work in the field of VNF placement and machine learning. Section
III outlines the methodology. Section IV describes the implementation
and discusses the obtained results. Finally, Section V discusses opportunities
for future work and concludes the paper.

\section{Related Work}

The following outlines some of the work being done in the field related
to the use of machine learning in VNF provisioning, allocation, and
function placement.

Khezri, \textit{et al.} \cite{key-7} propose a dynamic reliability-aware
NFV service provisioning solution that incorporates a deep Q-learning
network. Zhang \textit{et al.} \cite{key-8} propose an intelligent
cloud resource management framework that encompasses both deep and
reinforcement learning. Xu \textit{et al.} \cite{key-9} propose an
application-aware VNF deployed on a GPU server that analyzes packets
and classifies their application type using a deep neural network.
Sun, \textit{et al.} \cite{key-10} propose a dynamic SFC deployment
strategy using Q-learning. Riera \textit{et al.} \cite{key-11} suggest
the use of reinforcement learning to solve the service mapping problem.

All of the works presented above aim at solving the problem of VNF
resource allocation or placement using machine learning techniques.
The objective of each of the aforementioned papers often relates to
the constraint of delay, resource, and QoS guarantees or any combination
thereof. Our work differs from the works above as it considers the
minimization of delay across dependent VNF instances and the entire
SFC while simultaneously addressing performance and QoS requirements.

The previous work of Hawilo \textit{et al.} \cite{key-6} outlines
the use of a heuristic model to calculate the near-optimal placement
of VNF instances forming a service chain that minimizes the delay
between two dependent VNFs. Building off of their previous work, this
paper implements a machine learning model, which learns from the previous
placements performed by an adaptation of the aforementioned heuristic
in order to minimize the delay between future placements of dependent
VNF instances. Considering the delay between dependent components
is essential for maintaining service performance.

The major contributions of this work can be summarized as follows: 
\begin{lyxlist}{00.00.0000}
\item [{a)}] The implementation of a machine learning model to identify
the best placement of dependent VNF instances forming an SFC. 
\item [{b)}] The implementation of a machine learning model that learns
from previous effective placements when deciding the servers for future
placements. 
\item [{c)}] The implementation of a machine learning model that learns
operational constraints when selecting the best server that meets
QoS requirements.
\end{lyxlist}

\section{Methodology}

The following section will discuss the methodology behind our work
including the motivation, the generation of the dataset, and the advantages
of using a machine learning solution.

\subsection{QoS Requirements}

\begin{figure}[htbp]
\centerline{\includegraphics[width=7.5cm,height=7.5cm]{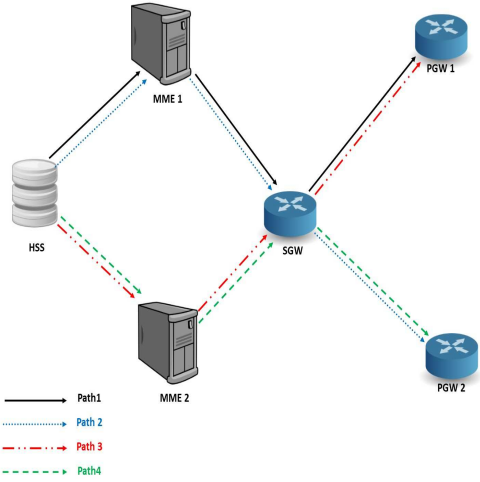}}
\caption{Computational Paths in Small-Scale Network}
\label{fig-2} 
\end{figure}

Usually, as a way of creating a more resilient system that is less
susceptible to failure, several instances of the same type are located
throughout the network. This introduces the concept of computational
paths \cite{key-6} which essentially indicates the number of different
routes network traffic can take to traverse an SFC. Fig.~\ref{fig-2}
illustrates this concept by showing various paths between interdependent
VNF instances forming an SFC. For a computational path to be successful,
the selected servers hosting the VNF instances should meet their computational
needs and ensure the delay tolerance between interdependent VNF instances
is not violated. However, the QoS requirements address much more than
simply the computational and delay requirements of the SFC. Further
consideration must be taken to include the availability and dependency
constraints that ensure the SFC is resilient to unexpected changes
in the network and traverses the VNFs in the correct order.

\subsection{Dataset Generation}

The dataset used to train the machine learning model is generated
by using an adaptation of the BACON algorithm presented by Hawilo
\textit{et al.} \cite{key-6} to place VNF instances. The algorithm
is selected for its ability to achieve near-optimal placement with
significantly decreased computational complexity compared to the Mixed
Integer Linear Programming model. The objective of the optimization
model is to minimize the delay between two dependent VNF instances
forming an SFC. The dataset contains placements from two network layouts.

The first network represents a small-scale network with 15 servers
and 6 VNF instances while the second network represents a medium-scale
network with 30 servers and 10 VNF instances to place. The parameters
of the network include server-to-server delay, server resources, VNF
instance resource requirements, and VNF instance delay tolerances.
They are all generated following the structure of a three-tier data
center. The instances are then placed using the aforementioned BACON
algorithm yielding the near-optimal placement \cite{key-6}. This
is conducted 10,000 times for each network, resulting in the placement
of the VNF instances given the respective trial's network conditions.
TABLE 1 lists the components of the two network models evaluated.

\begin{table}[htbp]
\caption{Network Components}

\centering{}%
\begin{tabular}{|c|c|c|}
\hline 
Component  & Network 1  & Network 2\tabularnewline
\hline 
\hline 
Total Available Servers  & 15  & 30\tabularnewline
\hline 
MME VNF Instance  & 2  & 2\tabularnewline
\hline 
HSS VNF Instance  & 1  & 3\tabularnewline
\hline 
SGW VNF Instance  & 1  & 3\tabularnewline
\hline 
PGW VNF Instance  & 2  & 2\tabularnewline
\hline 
\end{tabular}
\end{table}

\subsection{Advantages and Benefits of Machine Learning Based NFV Placement Algorithms}

In conventional, model-based networks, mathematical modelling is used
to describe the system's behaviour however, this approach is subject
to several limitations \cite{key-12}. Firstly, developing an all-encompassing
statistical network model has become challenging as network complexity
continually increases. Furthermore, NSPs are often faced with the
task of solving NP-Hard problems (ex. NFV Resource Allocation Problem,
VNF Placement Problem, etc.). The solution to these problems is computationally
expensive and often requires the use of sub-optimal heuristics to
achieve a solution in a feasible amount of time. Another concern with
model-based networks comes from the notion of problem decomposition
whereby a parent problem is split into smaller child problems which
are individually solved. Finding an optimal solution to a child problem
does not always inherently result in the optimal solution of the parent
problem \cite{key-12}.

As a method of mitigating the aforementioned limitations of model-based
networks, an alternate data-based framework leveraging the use of
machine learning has been suggested as a solution \cite{key-12}.
Data-based networking moves away from rigid mathematical modeling
and instead learns a model through the data generated from the network.
By learning from the data directly, several benefits arise. Firstly,
an exact mathematical model capturing every element of the system
is not required to describe the behaviour of the system as this is
inherently learned through the direct processing of the generated
network data. Additionally, as network complexity increases and changes
are made, the new data generated will provide the grounds for the
learning of the new system behaviour.

Perhaps the greatest advantage of the use of machine learning in data-based
networks is the reduction of the computational complexity of the system.
The operation of a machine learning algorithm can be divided into
the training and implementation phases; the training phase, containing
most of the computationally intense processes of the algorithm, can
be completed offline allowing for the implementation phase to be executed
during runtime. This is a major improvement over the use of complex
optimization models which are computationally intensive as their complexity
often scales with network parameters and is often inadequate during
runtime \cite{key-12}.

When considering the requirements of the incoming 5G network technology
(self-configuration, self-optimization, and self-healing), machine
learning is a natural candidate technology \cite{key-13}. Due to
the variety of algorithms available, machine learning can have a wide
range of functionalities in future networks (classification, forecasting,
clustering, etc.) to solve challenges currently faced by NSPs including
resource allocation, performance, security, resilience, energy efficiency,
and traffic management \cite{key-14}.

\subsection{Machine Learning Model}

The features extracted from the previous dataset generation stage
include resource requirements for the VNF instances, resource capacity
of the network servers, the delay between servers, delay tolerance
between dependent VNF instances, and component dependency. These metrics
are used to predict the placement of each of the VNF instances on
one of the network's servers. The machine learning model here can
be treated as a multi-output, multi-class classification problem as
there is a prediction for each of the VNF instances (multi-output)
and each prediction selects a server from the set of network servers
(multi-class).

Due to the nature of the problem, two approaches are evaluated, neighbour-based
algorithms and tree-based algorithms. These two families of algorithms
are selected for their ability to address the multi-class, multi-output
requirement, something which many learning algorithms are unable to
do due to the inherent complexity of the solution \cite{key-15}.
After training the model and performing a 10-fold cross-validation
test, it is determined that the tree-based algorithms are the best
performing family of algorithms specifically the decision tree algorithm.

The building of the decision tree follows a top-down approach starting
with the root node. The goal of the tree is to purify the nodes by
increasing the homogeneity of their associated samples. There are
two main metrics used to assess the purity of the node, Gini index
(1) and entropy (2) where the probability $p_{mk}$ is the proportion
of class observations at a given node \cite{key-15}. The proposed
decision tree uses the optimized Classification And Regression Tree
(CART) algorithm \cite{key-15}. This algorithm uses the Gini index
by default for evaluating node purity. 
\begin{equation}
Gini(X_{m})=\mathop{\sum_{i=1}^{k}p_{mk}(1-p_{mk})}
\end{equation}

\begin{equation}
H(X_{m})=\mathop{-\sum_{i=1}^{k}p_{mk}(\log_{2}p_{mk})}
\end{equation}
When initially constructing the tree, the Gini index of all features
with respect to the output label is calculated. The feature with the
lowest Gini index has the most purity and therefore is selected as
a root node. Once the root node is selected the dataset is split into
subsets depending on the feature values using thresholding. The Gini
index of each of the features in each respective subset is then calculated
and the lowest is selected as a branching attribute. This process
is completed until homogenous leaf nodes (pure) are achieved or the
maximum depth of the tree is reach. Since decision trees are prone
to overfitting when dealing with large quantities of data, the maximum
depth of the tree is set, essentially limiting the tree's ability
to expand vertically, and 10-fold cross-validation is used to further
ensure overfitting does not occur during the training phase.

By using the data generated from the previous placement of VNF instances
we create a data-driven network model whereby the algorithm is responsible
for determining and extracting the inherent relationships from the
data. Using this method, the complex mathematical modelling of the
system is bypassed, however, as demonstrated in the results section,
the algorithm has learned from the training phase and is able to predict
placement that approaches and outperforms the results obtained from
the heuristic.

\subsection{Time Complexity}

When comparing the computational complexity of the two methods, the
Delay Aware Tree (DAT) method exhibits a computational complexity
$O(n_{features}\cdot n_{samples}\cdot\log n_{samples})$ when creating
the tree and $O(\log n_{samples})$ when executing a query \cite{key-15}.
The original algorithm has a computational complexity of $O(\frac{s^{3}-s^{2}}{2})$
where s denotes the number of available servers in the network\cite{key-6}.
Comparing these two complexities, we can see that during runtime,
the DAT method would operate at a lower complexity since its initial
training phase would have already been completed and it would simply
need to execute the query request.

\section{Algorithm Performance Comparison}

In order to evaluate the results of the machine learning model, it
is compared to the BACON algorithm \cite{key-6}. The performance
is measured by calculating the delays between interconnected VNF instances
once the placement is made. Furthermore, the overall delay of each
computational path is also calculated.

\subsection{Implementation Setup}

The generation of the dataset is executed in Java while the data processing
and machine learning models are implemented using Python. Both the
generation of the dataset and the model implementation are run on
a PC with an Intel \textsuperscript{\textcopyright} Core\texttrademark{} i7-8700 CPU @ 3.20 GHz CPU,
32 GB RAM, and an NVIDIA GeForce GTX 1050 Ti GPU.

\subsection{Results}

\begin{figure}[h]
\centerline{\includegraphics[width=8cm,height=5cm]{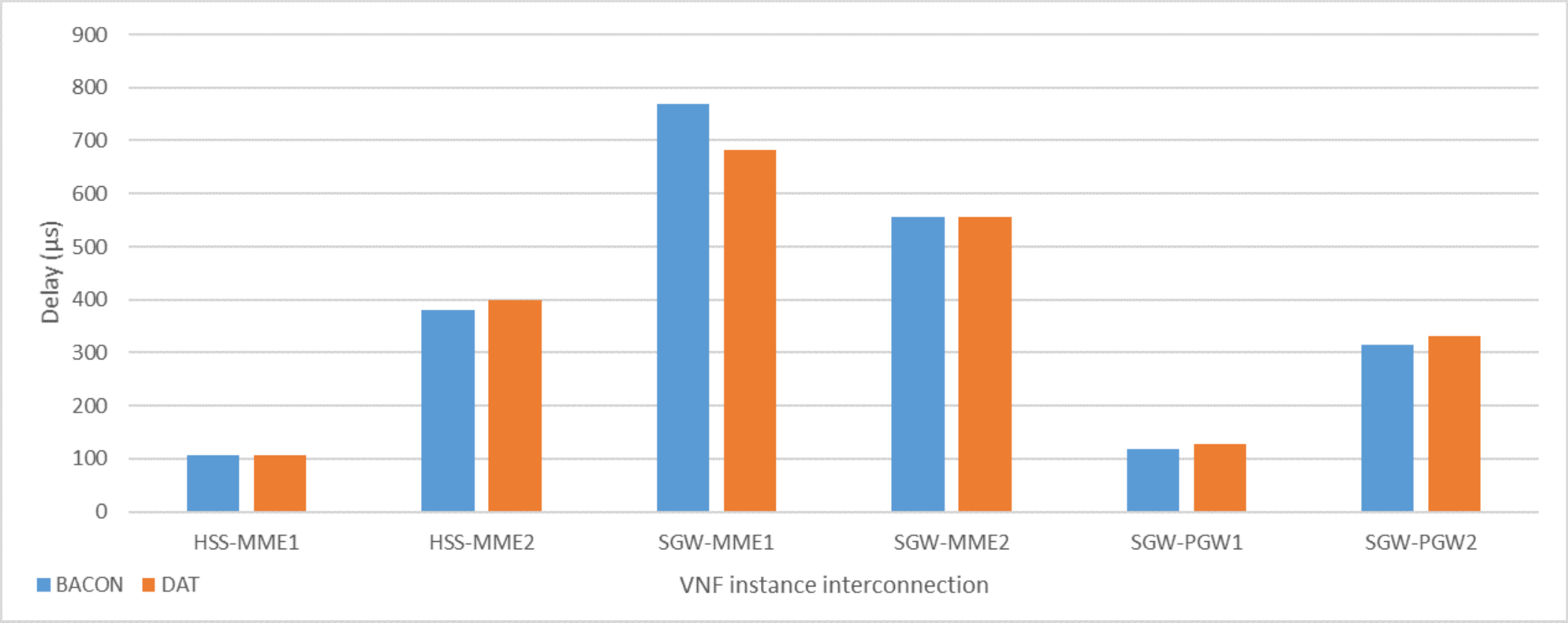}}
\caption{Delay between Interconnected VNF Instances}
\label{fig-3} 
\end{figure}

\begin{figure}[h]
\centerline{\includegraphics[width=7.5cm]{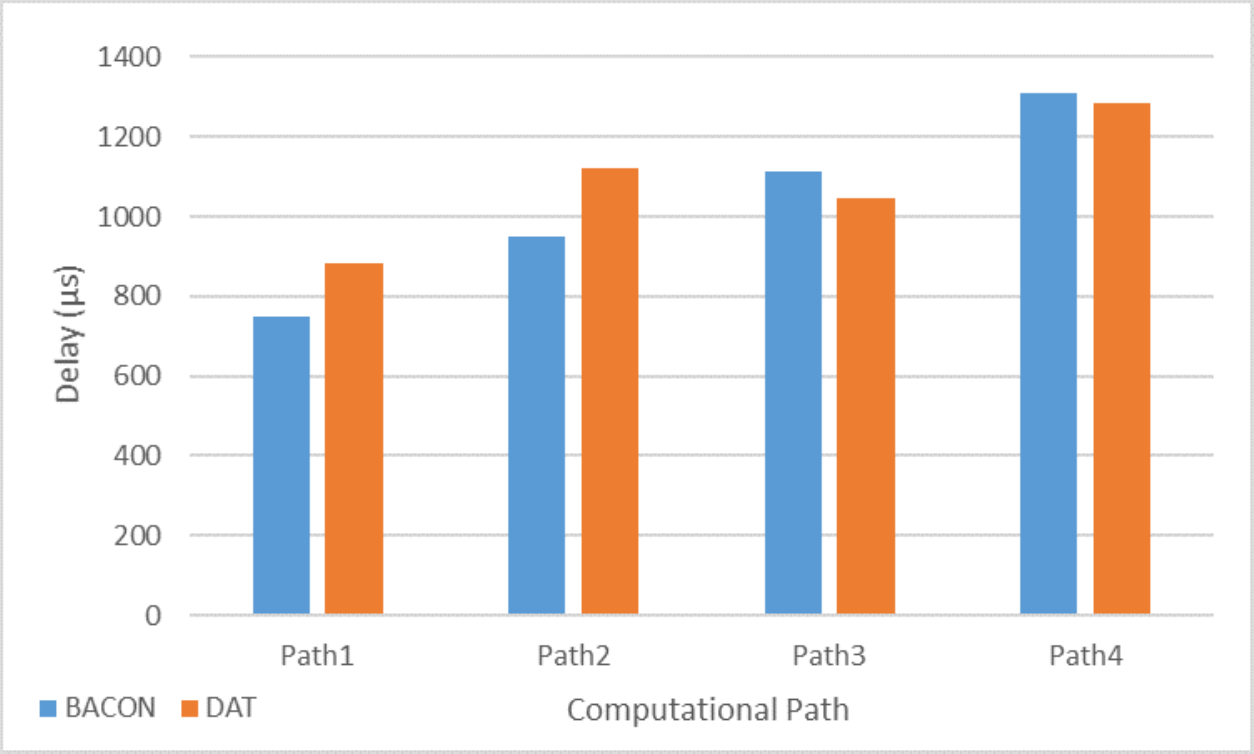}} \caption{SFC End-to-End Delay}
\label{fig-4} 
\end{figure}

\begin{figure}[h]
\centerline{\includegraphics[width=6cm]{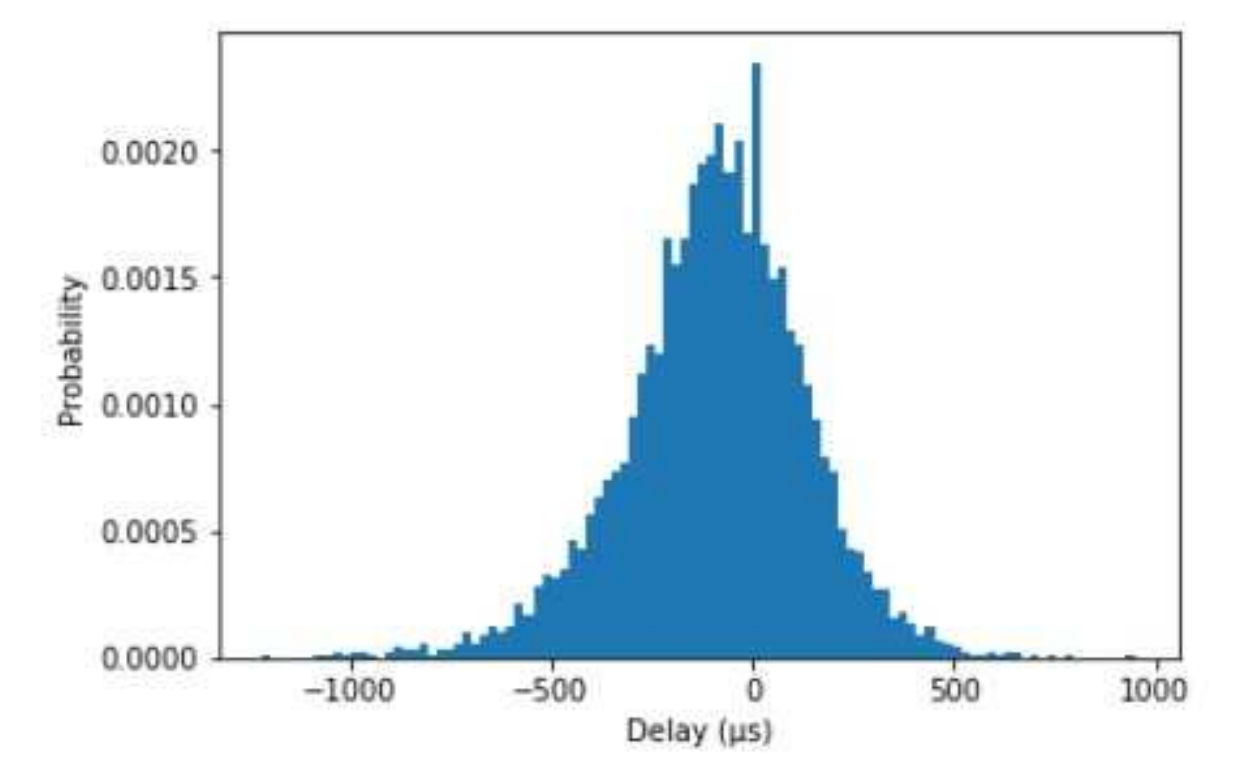}}
\caption{Delay Difference between BACON and DAT per Computational Path}
\label{fig-5} 
\end{figure}

The results of the simulations are displayed below. In terms of the
small scale 15 server network Fig.~\ref{fig-3} displays the delay
between the various interdependent VNF instances. Fig.~\ref{fig-4}
presents the overall SFC delay across the 4 computational paths previously
discussed. As seen from these results, it is clear that the Delay
Aware Tree (DAT) placement model has learned well from the near-optimal
placement of the BACON algorithm. When comparing the delay between
the interconnected VNF instances, DAT shows very good performance
as the delays observed are very close to the delays observed through
BACON's placement. In this case, it can be seen that DAT performs
slightly better than BACON for the first two paths and slightly worse
for the last two paths. Fig.~\ref{fig-5} shows the PDF of the difference
between the delay of the computational paths using BACON and DAT.
The mean of the distribution suggests that across all computational
paths, BACON has on average 34\textgreek{m}s less delay compared to
DAT. However, the left tail of the distribution greatly skews the
mean through outliers. By optimizing the model hyper-parameters the
overall performance of a machine learning algorithm improves\cite{key-16}.
By improving the performance of DAT through hyper-parameter optimization,
the mean of the distribution will shift further towards the positive
side and the tails of the distribution will be suppressed.

\begin{figure*}[t]
\centerline{\includegraphics[width=16cm,height=8cm]{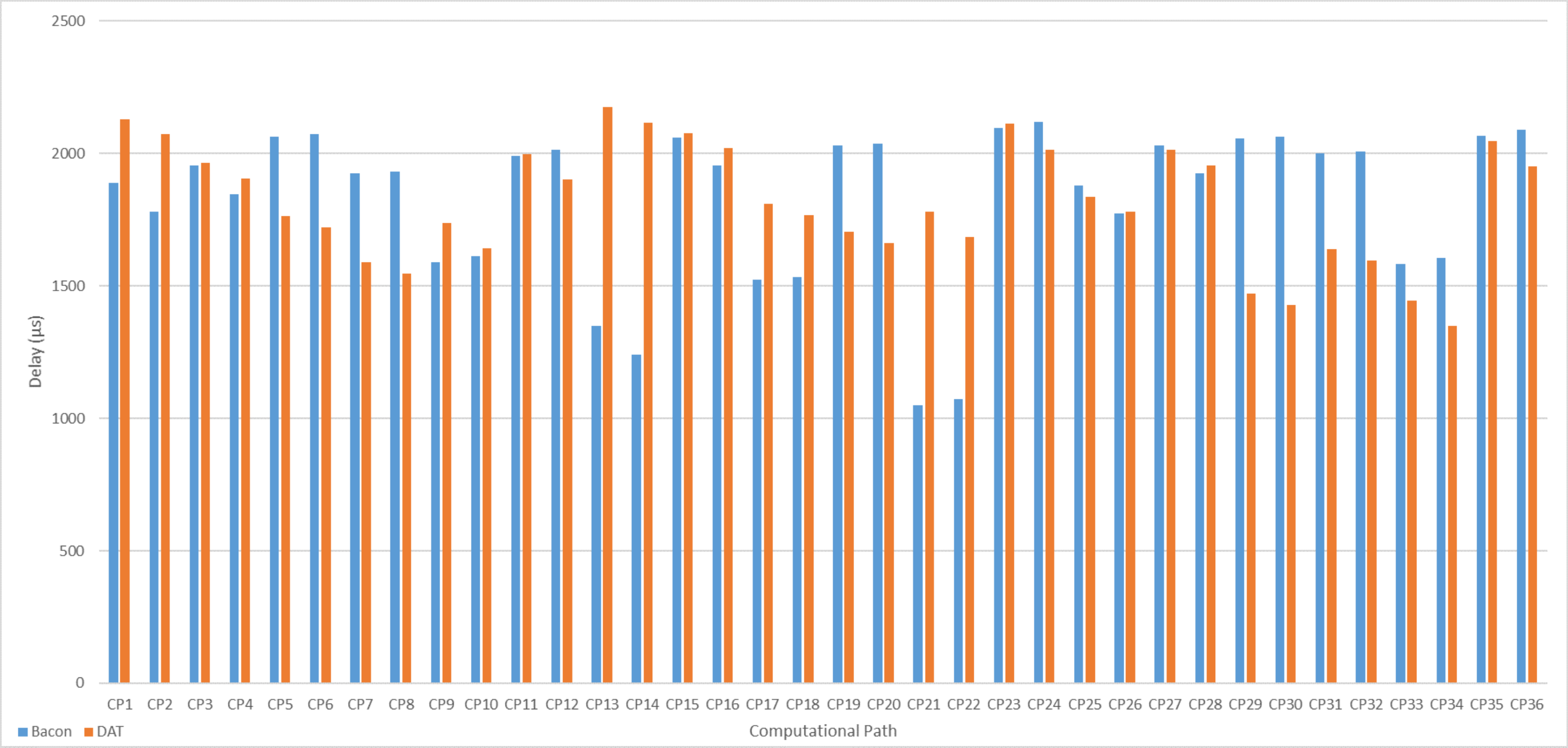}}
\caption{Computational Path Delay -- 30 Server Layout}
\label{fig-6} 
\end{figure*}

Fig.~\ref{fig-6} displays the delay across the 36 computational
paths in the medium 30 server network. As seen in the figure, DAT
continues to perform well despite the increase in network size. Assuming
a maximum allowable delay is imposed at 2000\textgreek{m}s we can
see that DAT successfully produces more computational paths that don't
violate this threshold thereby increasing the resiliency of the network.

\subsection{Hyper Parameter Optimization}

When considering the CART algorithm, there are four main hyperparameters:
tree depth, minimum sample split, minimum sample leaf, and the maximum
number of features considered. The work of Mantovani \textit{et al.}
\cite{key-17} deals with the tuning of decision tree hyperparameters
across several datasets. Results of their work suggest the min samples
split and the min samples leaf parameters are the most impactful on
a model's predictive success.

Given the set of hyperparameters $h$, the set of objectives $f$,
and the importance vector of each objective expressed through the
set of weights $w$, we can consolidate the multi-objective optimization
problem into a single objective function $O$ \cite{key-18}. The
evaluation criteria $E$ is an expression of the consolidated objective
function, the model trained with the set of hyperparameters $h$,
the training set $T_{s}$ and the validation set $V_{s}$ \cite{key-17}
Assuming $E$ is formulated as a cost function, the optimization problem
when implementing a k-fold cross-validation is the minimization of
$P(h)$.

\begin{equation}
h=\begin{cases}
h_{1},h_{2},\ldots,h_{n} & \}\end{cases}
\end{equation}

\begin{equation}
f=\begin{cases}
f_{1},f_{2},\ldots,f_{n} & \}\end{cases}
\end{equation}

\begin{equation}
w=\begin{cases}
w_{1},w_{2},\ldots,w_{n} & \}\end{cases}
\end{equation}

\begin{equation}
O=w_{1}f_{1}+w_{2}f_{2}+\ldots+w_{n}f_{n}
\end{equation}

\begin{equation}
E(O,model(h),T_{z},V_{s})
\end{equation}

\begin{equation}
P(h)=\frac{1}{k}\mathop{\sum_{i=1}^{k}}E(O,model(h),T_{z}^{(i)},V_{s}^{(i)})
\end{equation}

\[
Objective:
\]

\begin{equation}
minimize\;P(h)
\end{equation}

The hyperparameter optimization of the decision tree model was not
considered in this work; it will, however, be considered in future
work.

\section{Conclusion}

The work presented in this paper describes the first step towards
an implementable, intelligent, and delay-aware VNF placement strategy
for the NFV Orchestrator. Future work relates to the improvement of
the current model.

In terms of model improvement, hyperparameter optimization will occur
using the method previously outlined. When performing this optimization,
care must be taken to prevent over-fitting, something which plagues
large decision trees and hinders predictive accuracy. Furthermore,
a model will be built for a large data center network with a significant
increase in the number of available servers and VNF instances. Additionally,
the computational complexity of building the decision tree is dependent
on the number of features, dimensionality reduction techniques will
be used to improve the time required to train the model.

In order to deal with the rising costs of operating networks and the
new challenges facing NSP due to the rapid increase in connectivity
demand in data-dependent devices, NFV has been introduced as a solution.
By abstracting the network function from the underlying hardware,
we are able to move away from proprietary hardware and move towards
a more portable, flexible, and scalable network model. These benefits,
however, give rise to new challenges including the implementation
of a network model that can adhere to QoS requirements. The work presented
in the paper successfully implements and trains a delay-aware decision
tree model, DAT, which is able to learn the near-optimal placement
of VNF instances forming an SFC. When compared to the original placement
algorithm our model exhibits comparable performance and in certain
metrics outperforms the initial BACON algorithm.

\vspace{12pt}

\end{document}